\documentclass[12pt, letterpaper]{JHEP3}

\usepackage{amssymb, amsmath, amsopn, amsthm}

\usepackage{epsfig}

\newcommand{\eref}[1]{(\ref{#1})}

\newcommand{\cref}[1]{Chapter~\ref{#1}}
\newcommand{\beq}{\begin{equation}}
\newcommand{\eeq}{\end{equation}}
\newcommand{\ba}{\begin{array}}
\newcommand{\ea}{\end{array}}
\newcommand{\bcenter}{\begin{center}}
\newcommand{\ecenter}{\end{center}}

\def\IB{\relax\hbox{$\inbar\kern-.3em{\rm B}$}}
\def\IC{\relax\hbox{$\inbar\kern-.3em{\rm C}$}}
\def\ID{\relax\hbox{$\inbar\kern-.3em{\rm D}$}}
\def\IE{\relax\hbox{$\inbar\kern-.3em{\rm E}$}}
\def\IF{\relax\hbox{$\inbar\kern-.3em{\rm F}$}}
\def\IG{\relax\hbox{$\inbar\kern-.3em{\rm G}$}}
\def\IGa{\relax\hbox{${\rm I}\kern-.18em\Gamma$}}
\def\IH{\relax{\rm I\kern-.18em H}}
\def\IK{\relax{\rm I\kern-.18em K}}
\def\IL{\relax{\rm I\kern-.18em L}}
\def\IP{\relax{\rm I\kern-.18em P}}
\def\IR{\relax{\rm I\kern-.18em R}}
\def\IZ{\relax\ifmmode\mathchoice
{\hbox{\cmss Z\kern-.4em Z}}{\hbox{\cmss Z\kern-.4em Z}}
{\lower.9pt\hbox{\cmsss Z\kern-.4em Z}}
{\lower1.2pt\hbox{\cmsss Z\kern-.4em Z}}\else{\cmss Z\kern-.4em Z}\fi}
\def\II{\relax{\rm I\kern-.18em I}}

\def\sCC{{\kern 0.27em\vrule height1.45ex width0.03em depth0em
          \kern-0.30em\rm C}}
\def\C{{\mathchoice
  {\sCC}
  {\sCC}
  {\kern 0.225em \vrule height1.05ex width0.025em depth0em \kern-0.25em \rm C}
  {\kern 0.180em \vrule height0.78ex width0.02em depth0em \kern-0.2em \rm C}
        }}
\def\sHH{{\rm I\kern-.16em{}H}}
\def\H{{\mathchoice
  {\sHH}
  {\sHH}
  {\rm I\kern-.13em{}H}
  {\rm I\kern-.13em{}H} }}
\def\sNN{{\rm I\kern-.16em{}N}}
\def\N{{\mathchoice
  {\sNN}
  {\sNN}
  {\rm I\kern-.12em{}N}
  {\rm I\kern-.10em{}N} }}
\def\sPP{{\rm I\kern-.16em{}P}}
\def\P{{\mathchoice
  {\sPP}
  {\sPP}
  {\rm I\kern-.12em{}P}
  {\rm I\kern-.10em{}P} }}
\def\sQQ{{\kern 0.27em \vrule height1.45ex width0.03em depth0em
          \kern-0.30em \rm Q}}
\def\Q{{\mathchoice
        {\sQQ}
        {\sQQ}
  {\kern 0.225em \vrule height1.05ex width0.025em depth0em \kern-0.25em \rm Q}
  {\kern 0.180em \vrule height0.78ex width0.020em depth0em \kern-0.20em \rm Q}
        }}
\def\sRR{{\rm I\kern-0.16em{}R}}
\def\R{{\mathchoice
  {\sRR}
  {\sRR}
  {\rm I\kern-0.12em{}R}
  {\rm I\kern-0.10em{}R} }}
\def\sZZ{{\rm Z\kern-0.32em{}Z}}
\def\Z{{\mathchoice
  {\sZZ}
  {\sZZ} 
  {\rm Z\kern-0.3em{}Z}     
  {\rm Z\kern-0.25em{}Z} }}  
\def\ZZZ{{\rm Z\kern-0.24em{}Z}}
\def\sII{{\rm I\kern-0.16em{}I}}
\def\I{{\mathchoice
  {\sII}
  {\sII}
  {\rm I\kern-0.12em{}I}
  {\rm I\kern-0.10em{}I} }}

\def\Tr{{\rm Tr}}

\def\inbar{\,\vrule height1.5ex width.4pt depth0pt}
\font\cmss=cmss10 \font\cmsss=cmss10 at 7pt

\def\smiley{\hbox{\large$\bigcirc$\hspace{-0.80em}\raise.2ex
\hbox{$\cdot\cdot$}\kern-.61em\lower.2ex\hbox{\scriptsize$\smile$}}\ }
\def\frowny{\hbox{\large$\bigcirc$\hspace{-0.80em}\raise.2ex
\hbox{$\cdot\cdot$}\kern-.635em\lower.2ex\hbox{\scriptsize$\frown$}}\ }

\def\I{{\rlap{1} \hskip 1.6pt \hbox{1}}}

\makeatletter
\let\hangafter\@hangfrom
\makeatother

\def\makeatletter{\catcode`\@=11}
\makeatletter
\def\mathbox#1{\hbox{$\m@th#1$}}%
\def\math@ccstyles#1#2#3#4#5#6#7{{\leavevmode
      \setbox0\mathbox{#6#7}%
      \setbox2\mathbox{#4#5}%
      \dimen@ #3%
      \baselineskip\z@\lineskiplimit#1\lineskip\z@
      \vbox{\ialign{##\crcr
             \hfil \kern #2\box2 \hfil\crcr
             \noalign{\kern\dimen@}%
             \hfil\box0\hfil\crcr}}}}
\def\mathaccstyles{\math@ccstyles\maxdimen}
\def\maththroughstyles{\math@ccstyles{-\maxdimen}}
\def\unity%
 {\maththroughstyles{.45\ht0}\z@\displaystyle {\mathchar"006C}\displaystyle 1}

\title{Single-Sector Supersymmetry Breaking in Supersymmetric QCD}

\author{Sebasti\'an Franco$^1$ and Shamit Kachru$^{1,2,3}$

\\

~\\

$^1$Kavli Institute for Theoretical Physics, University of California \\
Santa Barbara, CA 93106\\
 \vspace{0.3cm}

$^2$Department of Physics, University of California \\
Santa Barbara, CA 93106\\
\vspace{0.3cm}

$^3$Department of Physics and SLAC, Stanford University\\
Stanford, CA 94305\\

\vspace{0.8cm}

\email{sfranco@kitp.ucsb.edu, skachru@kitp.ucsb.edu}\\

}

\abstract{
We construct examples of single-sector supersymmetry breaking based on simple deformations of supersymmetric QCD with (weakly) gauged flavor group. These theories are calculable in a weakly coupled Seiberg dual description. In these models, some of the particles in the first two generations of quarks and leptons are composites of the same strong dynamics which
leads to dynamical supersymmetry breaking. Such models can explain the hierarchies of 
Yukawa couplings in the Standard Model, in a way that predictively correlates with the spectrum
of SUSY-breaking soft terms.  

}

\preprint{NSF-KITP-09-114 \\ SLAC-PUB-13717 \\ SU-ITP-09/35}

\def\be{\begin{equation}}

\def\ee{\end{equation}}

\def\bea{\begin{eqnarray}}

\def\eea{\end{eqnarray}}

\begin{document}

\tableofcontents

\section{Introduction}

Realistic supersymmetric models usually assume a hidden sector where supersymmetry breaks, a visible sector containing the MSSM or some extension thereof, and
a messenger sector which communicates the SUSY-breaking to the MSSM, generating a 
predictive pattern of soft terms.  In models of gravity mediation, this communication typically occurs
through moduli-like fields which introduce direct Planck-suppressed couplings between the
hidden sector F-terms and the MSSM fields (through dimension six terms in the K\"ahler potential).  
In gauge mediation, instead the messengers are
massive fields carrying Standard Model gauge charges, which themselves may feel supersymmetry
breaking in a variety of different ways.   Loop effects controlled by Standard Model gauge
couplings then transmit the SUSY-breaking to the Standard Model fields.  

It is natural to ask if there are simpler models which have a less modular structure.  In this paper, we 
explore an idea suggested in \cite{ALT,LT}.   These authors suggested that there is a single
sector which both breaks supersymmetry and produces (some of) the quarks and leptons of
the Standard Model as composites of the same strong dynamics.  They produced a variety of
examples which seem to accomplish this basic goal.  However, the existing examples are not
calculable -- there is no small parameter in terms of which the precise vacuum structure or 
soft-terms can be computed.
In this paper, we pursue the construction of calculable examples of this basic idea.  We will find that such models are easily realized in simple deformations of supersymmetric QCD (which include extra gauge singlet ``spectators" to the QCD dynamics).  The compositeness
scale will be the strong-coupling scale of the QCD theory, which we typically take to be a high
scale, while the SUSY-breaking dynamics will occur at parametrically lower energies (giving rise
to ${\rm TeV}$-scale soft terms).

The great attraction of this idea of single-sector SUSY breaking is that, in addition to avoiding the modular structure of
gravity and gauge mediation, it correlates the SUSY-breaking soft terms in a predictive way with
a possible model of flavor physics.  The composite generations naturally arise from products of basic fields in the high energy theory; their couplings to the Higgs are then secretly
higher-dimension operators, suppressed by a flavor scale $M_{\rm flavor}$.   Therefore, the composite
generations naturally have suppressed Yukawa couplings.  This suggests that one should
construct models where the first two generations are composite, while the third generation 
is a spectator to the strongly-coupled dynamics.  Then it naturally transpires that the
soft mass of the top squark will be parametrically smaller than the soft masses of the
first two generation squarks, for instance; the latter directly feel the strongly coupled SUSY
breaking dynamics, while the stop only gets a mass through gauge mediation effects feeding
down from the higher scale soft masses.  In this way, one obtains a predictive structure where
the flavor properties of the Standard Model are directly reflected also in the spectrum of
soft masses.    In fact, such a spectrum of soft terms was proposed as being particularly phenomenologically 
attractive in earlier work \cite{Dine,Giudice,Pomarol,Nelson}, even without the simple 
dynamical explanation provided by this class of models.

It is important to be sure that any direct coupling of the Standard Model fields to the SUSY-breaking sector does not generate the soft masses at tree level through renormalizable interactions.  In that case, the Dimopoulos-Georgi theorem \cite{DG}
(which follows from a simple examination of supertraces of mass-matrices arising from tree level
SUSY-breaking) would guarantee the existence of a squark which is lighter than one of the first
generation quarks.   The strongly-coupled single-sector models can evade this theorem as in
\cite{ALT,LT}.  In our theories, where the supersymmetry breaking dynamics can be analyzed
perturbatively in a magnetic dual description as in \cite{ISS}, we will see that the theorem is
evaded because the soft terms are generated by loop effects in the magnetic theory.  

The organization of this paper is as follows.
In \S2, we review the existence of SUSY-breaking vacua in supersymmetric QCD (SQCD).  In \S3,
we describe various modifications we make to this theory to produce a single-sector model.
We focus on the case that produces either one or two composite ${\bf \bar 5}$s of $SU(5)$.
In \S4, we describe various extensions that can produce up to two full composite
generations in the ${\bf 10} + {\bf \bar 5}$ of $SU(5)$.  While the models of \S3 and \S4 
do not have many extra matter fields at the TeV scale, this is because we push a large
number of extra charged fields to a high mass scale.  This is inelegant, so in \S5,
we also present some models where the electric gauge theory is slightly more complicated,
but where the number of extra multiplets (that would appear at a very high scale) is somewhat
ameliorated.
We close in \S6\ with some basic comments about the phenomenology of these models, and other
model-building ideas.
In the appendix, we show that various perturbations we make to the basic SUSY-breaking
metastable vacua in SQCD \cite{ISS} (to allow greatly improved phenomenology) do not
destabilize the SUSY-breaking vacuum.

\section{SUSY-breaking vacua in SUSY QCD}

We now provide a brief review of the metastable vacua of SQCD \cite{ISS}. This model is usually referred to as ISS. In the next section we find that simple examples of these theories can be extended 
to give rise to composite models.

The basic model of \cite{ISS} corresponds to SQCD in the free magnetic range $N_c+1 \leq N_f < {3\over 2} N_c$ with massive flavors 

\beq
W\, = \, m\, \Tr \, {\tilde Q} Q.
\label{W_electric}
\eeq

This theory has a dual description which captures the same infrared (IR) physics \cite{SeibergDual}.
The IR free magnetic dual has $SU(N)$ gauge group (where $N=N_f-N_c$), $N_f$ flavors $q$ and ${\tilde q}$, and mesons $M$. From the perspective of the electric theory, the mesons are composite operators: $M_{ij}=Q_i \cdot \tilde{Q}_j$.

The superpotential for the dual theory is
\beq
W\, =\, \frac{1}{\hat\Lambda} \, \Tr M q {\tilde q} \, +\, m\, \Tr M
\label{W_mag_1}
\eeq
where the last term follows from \eref{W_electric}. $\hat \Lambda$ is determined in terms of the electric and magnetic dynamical scales $\Lambda_e$ and $\Lambda_m$
\beq
\Lambda_e^{\, 3N_c-N_f}\, \Lambda_m^{3(N_f-N_c)-N_f}\, =\, 
\hat \Lambda^{N_f}
\label{match}
\eeq

We can define rescaled mesons with canonical dimension $\Phi= M/\Lambda_e$, and the couplings $h=\Lambda_e/{\hat\Lambda}$ and 
$\mu^2=-m\hat\Lambda$. Then, \eref{W_mag_1} becomes
\beq
W\, =\, h \, \Tr\, q\, \Phi\, {\tilde q}\, -\, h\mu^2 \Tr \, \Phi
\label{W_mag_2}
\eeq

For $\mu=0$, the global symmetry (recall that the $SU(N)$ gauge group is IR free and hence becomes global) is 
\beq
SU(N) \times SU(N_f)^2 \times U(1)_B \times U(1)' \times U(1)_R 
\label{global_1}
\eeq
The fields in the magnetic theory $q$, $\tilde{q}$ and $\Phi$ transform as $({\bf N},{\bf \overline{N}_f},{\bf 1})$, $({\bf \overline{N}},{\bf 1},{\bf N_f})$ and $({\bf 1},{\bf N_f},{\bf \overline{N}_f})$ under $SU(N)\times SU(N_f)\times SU(N_f)$. For $\mu\neq 0$, \eref{W_mag_2} breaks the global symmetry group down to
\beq
SU(N) \times SU(N_f) \times U(1)_B \times U(1)_R 
\label{global_2}
\eeq
with $SU(N_f)$ the diagonal subgroup in $SU(N_f)^2$. We can parametrize fields as

\beq
q^T = \left(\begin{array}{c} \chi_{N \times N}\\
\rho_{N_c\times N}\end{array}\right),\ \ \ \ \tilde q = \left(\begin{array}{c} \tilde \chi_{N \times N}\\
\tilde \rho_{N_c\times N}\end{array}\right) \ \ \ \ \Phi=\left(\begin{array}{ccc} Y_{N \times N} & & Z^T_{N \times N_c} \\
\tilde{Z}_{N_c \times N} & & \Phi_{0,N_c \times N_c} \end{array}\right)\, .
\label{field_parametrization}
\eeq

Supersymmetry is broken at tree-level by the rank-condition. There is a classical moduli space of non-supersymmetric vacua with vacuum energy $V_0=N_c |h^2 \mu^4|$, corresponding to $\langle \chi \tilde \chi \rangle = \mu^2$ and $\Phi_0$ arbitrary. The pseudomoduli, namely the classically flat directions that are not Nambu-Goldstone bosons of any broken global symmetry, are stabilized at 1-loop at

\beq
\Phi_0=0 \ \ \ \ \ \ \ \ \chi = \tilde \chi = \mu \, {\bf 1}_{N}
\label{metastable_vacuum}
\eeq
getting $\mathcal{O}(|h^2 \mu|)$ masses. All other directions are fixed at zero at tree level, with $\mathcal{O}(|h \mu|)$ masses. The theory also has a supersymmetric vacuum. In the limit $\epsilon \equiv \mu/\Lambda_m \sim \sqrt{m/\Lambda_e} \to 0$, the metastable vacuum is far from the supersymmetric one and is parametrically long lived .

The vev of the dual quarks in \eref{metastable_vacuum} breaks \eref{global_2} down to

\beq
SU(N)_D \times SU(N_c) \times U(1)_B \times U(1)_R 
\label{global_3}
\eeq
where we have used $N_f-N=N_c$.

Our goal is to construct models in which some of the MSSM fields are composite from the perspective of the UV theory. Hence, they should arise as (part of) the magnetic mesons $\Phi$. Furthermore, in order to circumvent the Dimopoulos-Georgi theorem, we want their soft masses not to be generated at tree level. As a result, the composites are constrained to live inside the pseudomoduli matrix $\Phi_0$.

\section{The simplest models: composite ${\bf \bar 5}$s of $SU(5)$}

We begin by writing down simple theories that generate just the ${\bf \bar 5}$ components of
one or two generations as composites.    Throughout the paper we will abuse notation by referring to MSSM matter fields in terms of $SU(5)$ quantum numbers.  In part this is because our models are
naturally viewed as unified theories, and in part this is because it is more cumbersome to tabulate
the $SU(3) \times SU(2) \times U(1)$ quantum numbers individually; it is easy to go back and forth,
of course.

\subsection{A single ${\bf \bar 5}$}
We start with SQCD with $SU(6)$ gauge
group and $N_f = 7$ flavors of quarks.  
The $SU(7)_{\rm flavor}$ flavor symmetry will be higgsed down to $SU(6)_{\rm flavor}$ in the ISS vacuum, as 
described in \S2.
In terms of representations of the $SU(5)$ subgroup of
$SU(6)_{\rm flavor}$ which we are going to eventually (weakly) gauge, the quarks decompose as

\beq
\begin{array}{rcl}
Q & = & ({\bf 5} + {\bf 1}) + {\bf 1}\\
\tilde{Q} & = & ({\bf \overline{5}} + {\bf 1}) + {\bf 1}
\end{array}
\label{embedding_7}
\eeq
The parentheses group the $SU(5)$ representations in a given $SU(6)_{\rm flavor}$ representation.

The fields in the magnetic theory can be decomposed following \eref{field_parametrization}

\beq
q^T = \left(\begin{array}{c} \chi_{1 \times 1}\\
\rho_{6\times 1}\end{array}\right),\ \ \ \ \tilde q = \left(\begin{array}{c} \tilde \chi_{1 \times 1}\\
\tilde \rho_{6\times 1}\end{array}\right), \ \ \ \ \Phi=\left(\begin{array}{ccc} Y_{1 \times 1} & & Z^T_{1 \times 6} \\
\tilde{Z}_{6 \times 1} & & \Phi_{0,6 \times 6} \end{array}\right)\, ,
\eeq
with $Y$, $Z$, ${\tilde Z}$ and $\Phi_0$ transforming in the ${\bf 1}$, ${\bf \bar 6}$, ${\bf 6}$ and $(\rm{{\bf Adj}}+{\bf 1})$ of $SU(6)_{\rm flavor}$.

Since $\Phi \sim Q \cdot \tilde{Q}$, it follows from \eref{embedding_7} that $\Phi_0$ decomposes into

\beq
\Phi_0 = {\bf \bar 5} + \left[ \bf{24} + \bf{5} + \bf{1} +  \bf{1}\right],
\label{Phi_0_decomposition_1}
\eeq
where the ${\bar 5}$ which we wish to identify with the first generation
down-type quarks and $(e, \nu)$ appears outside of the brackets.  In addition to this matter
content, we introduce the further matter required to complete 3 full generations in the
$({\bf {10} + {\bf \bar 5}})$ of $SU(5)$ -- a ${\bf 10}$ we can call $\Psi^{(1)}$, and two $({\bf{10} + {\bf \bar 5}}$)s we
can denote by $\Psi^{(2,3)}$.  We additionally introduce  matter fields $X_{\bf \bar 5}$ and $ X_{\bf 24},$
in the
${\bf \bar 5}$ and ${\bf 24}$ representations.  We will pair these up with the extra unwanted $SU(5)$-charged
matter in (\ref{Phi_0_decomposition_1}), giving the extra multiplets a mass of order $\Lambda_{e}$.
Finally we imagine the standard Higgs content, with elementary Higgs scalars; we will omit further mention of these multiplets (while discussing the elementary MSSM generations only so it is completely clear which matter fields
are composite).

Concretely, in order to give masses to the unwanted matter in the mesons, we
add the following terms to the superpotential in the electric theory:

\begin{equation}
W_{3} = \lambda \left[ (Q\tilde Q)_{\bf 5} X_{\bf \bar 5} + (Q\tilde Q)_{\bf 24} X_{\bf 24} \right]~. 
\label{crapmass_1}
\end{equation}
These terms turn into mass terms in the magnetic dual, giving masses of ${\cal O}(\Lambda_{e})$
to the extra states for $\lambda \sim {\cal O}(1)$ (recall that one must re-scale the meson by a 
power of $\Lambda_{e}$ to canonically normalize it, in the magnetic theory).   
We will generally want to choose $\Lambda_{e}$ to be a very high energy scale, approaching
$M_{\rm GUT}$. 
Note that we are guaranteed that the chiral
fermion in the ${\bf \bar 5}$
of $SU(5)$ that arises from $\Phi_0$ will remain massless at the level of our 
discussion -- chirality in absence of $SU(5)$ breaking forbids it from obtaining a mass.

Finally, we need to worry about breaking the $U(1)_R$ which is unbroken in the metastable
SUSY-breaking vacuum.  Such an unbroken $U(1)_R$ would forbid (Majorana) gaugino masses.
The simplest way to break the $U(1)_R$ is to add a quartic superpotential perturbation in
the electric theory:
\begin{equation}
W_{4} = h^2 \mu_{\phi}  ~{\rm tr} ((Q \tilde Q)^2) ~.
\label{quartic}
\end{equation}

The perturbation in (\ref{quartic}), in absence of our $W_3$, has 
been studied extensively in the recent paper
\cite{Strassler}, building on the work of \cite{Torroba, Kutasov}. As in the $\mu_\phi=0$ case, the global symmetry is given by \eref{global_3}. This perturbation has several
important effects:

\bigskip

\noindent
$\bullet$ For $\mu_{\phi} \neq 0$, the quartic terms result in a tree level breaking of the $U(1)_R$ 
symmetry present in the unperturbed ISS vacuum.   It also results in a parametrically larger 
${\it spontaneous}$ breaking -- the $SU(6)_{\rm flavor}$ singlet ${\rm tr} \, \Phi_0$ acquires a
vacuum expectation value.

\bigskip
\noindent
$\bullet$
The coupling (\ref{quartic}) gives a small mass term pairing the ${\bf \bar 5}$ in $\Phi_0$ and the 
${\bf{5}}$ in $\Phi_0$.  This is non-zero but far smaller than the mass pairing the external $X_{\bf \bar 5}$ field to the ${\bf 5}$ in $\Phi_0$.  Therefore, the massless ${\bf \bar 5}$ is no longer precisely the composite
${\bf \bar 5}$ in $\Phi_0$ - there is some admixture of the external ${\bf \bar 5}$ $X_{\bf \bar 5}$.  However,
in the regime $h \sim 1,  \mu_{\phi} \ll \Lambda_e$, where we work, the massless ${\bf \bar 5}$ is almost
entirely the composite from $\Phi_0$.   A similar remark will apply in all other models in this paper.

\bigskip

\noindent
$\bullet$ For $\mu_{\phi} \ll \mu$, the 1-loop effective potential still has a reliable supersymmetry-breaking minimum. The vev of ${\rm tr} \, \Phi_0$ and the MSSM gaugino mass are $\sim \mu_\phi$.

\bigskip

\noindent
$\bullet$  We note that in \cite{Strassler}, an additional quartic double-trace perturbation
played an important role in giving masses to the charged fermions in the $\Phi_0$ supermultiplet.
In our case, the cubic perturbation (\ref{crapmass_1}) gives these particles a much larger mass,
and we do not need to activate the double-trace perturbation.
The fact that these extra states get a mass of ${\cal O}(\Lambda_e)$ in our
construction  is crucial for
avoiding Landau poles at relatively low energies.  Even if one is interested in just building
models of direct gauge mediation without composite quarks and leptons, it seems that 
introducing the extra spectators $X$ and pairing them up with the extra states in $\Phi_0$ as
in (\ref{crapmass_1}) may be a simple way to proceed to avoid Landau poles (or push them
up to much higher energies).

\bigskip

The reader may wonder if the presence of the 1-loop minimum discussed in 
\cite{Strassler} is robust in the presence of the cubic perturbations
(\ref{crapmass_1}).  We show that this is so in appendix A; it is important that in perturbing
the theory with (\ref{crapmass_1}), we did not add new couplings of the singlet
${\rm Tr}(\Phi_0)$.

In summary, the matter content in the electric theory is:

\beq
\begin{array}{c|ccc}
 & \ \ \ SU(6)_{\rm gauge} \ \ \ & & SU(5)_{\rm SM} \subset SU(6)_{\rm flavor} \\ \hline
\ \ \ Q \ \ \ & {\bf 6} & & {\bf 5} + {\bf 1} + {\bf 1} \\
\ \ \ \tilde{Q} \ \ \ & {\bf \overline{6}} & & {\bf \bar 5} + {\bf 1} + {\bf 1}\\
\Psi^{(1)} & {\bf 1} && {\bf 10} \\
\Psi^{(2,3)} & {\bf 1} & & {\bf 10}+{\bf \bar{5}} \\
X_{\bf \bar 5}& {\bf 1} & & {\bf \bar{5}}\\
X_{\bf  24}&{\bf 1}& &{\bf 24}
\end{array}
\label{matter_electric_1}
\eeq

The matter content in the magnetic theory is:
\beq
\begin{array}{cc}

\begin{array}{c}
\vspace{.5cm}\Phi \ \ \left\{\begin{array}{c}  \\ \\ \\ \\ \end{array}\right.
\end{array} &

\begin{array}{c|c}
 & \ \ \ SU(5)_{\rm SM} \subset SU(6)_{\rm flavor} \ \ \ \\ \hline
\ \ \ q \ \ \ & {\bf \overline{5}} + {\bf 1} + {\bf 1} \\
\ \ \ \tilde{q} \ \ \ &  {\bf 5} + {\bf 1}  + {\bf 1}\\ \hline
Y & {\bf 1} \\
Z & {\bf \overline{5}} + {\bf 1} \\
\tilde{Z} & {\bf 5} + {\bf 1} \\
\Phi_0 & {\bf \bar 5} + \left[ \bf{24} + \bf{5} + 2 \times \bf{1}\right] \\ \hline
\Psi^{(1)} & {\bf 10} \\
\Psi^{(2,3)} & {\bf 10}+{\bf \bar{5}} \\
X_{\bf \bar 5} & {\bf \bar{5}}\\
X_{\bf 24}&{\bf 24}
\end{array}
\end{array}
\label{matter_magnetic_1}
\eeq

The most rudimentary phenomenology of this class of models (with a focus on the that of the models with two full composite generations) will be briefly discussed in \S6.  The important feature in this
particular model is that
below the scale $\Lambda_{e}$ (which serves as the compositeness scale in these models), the only light $SU(5)$ charges are the three MSSM generations,
and vector-like ``messenger'' matter arising from the $\rho$, $\tilde \rho$, $Z$ and $\tilde Z$ fields. We refer to these fields as the $(\rho,Z)$ sector which, in this model, consists of two $({\bf 5}+{\bf \bar 5})$ pairs at the tree level mass scale $\sim h \mu$. More precisely, $2 N_c N$ real scalars in the $(\rho,Z)$ sector are Nambu-Goldstone bosons of the $SU(N_f)\to SU(N) \times SU(N_c)$ breaking of the global symmetry. These scalars become pseudo-Goldstone bosons once the MSSM gauge group is higgsed, picking a mass $\sim g_{SM} \mu/(4 \pi)$ \cite{Strassler}. This fact holds for the rest of our examples. The resulting theory will be a hybrid of a single-sector model and a model of direct
gauge mediation, where gauge mediated masses arise both from the composite Standard Model ${\bf \bar 5}$ and the $(\rho,Z)$ messengers.

\subsection{Two ${\bf \bar 5}$s}

One can obtain models with two composite ${\bf \bar 5}$s by a simple generalization of
the theory in \S3.1.  For the electric gauge theory, we choose $SU(7)$ gauge theory with
$N_f = 8$ flavors of quarks. The analysis proceeds by precise analogy to \S3.1, with 

\beq
\begin{array}{rcl}
Q & = & ({\bf 5}+ {\bf 1} + {\bf 1}) + {\bf 1}\\
\tilde{Q} & = & ({\bf \overline{5}} + {\bf 1}+ {\bf 1}) + {\bf 1}
\end{array}
\label{embedding_8}
\eeq
In terms of $SU(5)$ representations, ${\bf \Phi_0}$ decomposes into
\beq
{\bf \Phi_0} = {\bf \bar 5} + {\bf \bar 5} + \left[ {\bf 24} + {\bf 5} + {\bf 5} + 5  \times {\bf 1} \right]~.
\label{phimess}
\eeq
We introduce spectators $X_{\bf R}$ in the appropriate representations to pair up with
the unwanted matter in brackets in (\ref{phimess}), via a coupling completely analogous
to (\ref{crapmass_1}).  This gives these fields a mass of order
$\Lambda_{e}$.  We also add $SU(5)$ multiplets $\Psi^{(1,2)}$ in the ${\bf 10}$ and
$\Psi^{(3)}$ in the $({\bf 10} +{\bf \bar 5})$ to fill out the MSSM matter content.

In addition to the three generations of quarks and leptons (with the first two being partially
composite), we find two $({\bf 5} + {\bf \bar 5})$ ``messenger'' pairs from the $(\rho,Z)$ sector.

\section{Models with fully composite generations}

The models of \S3 are good as warmups, but would leave us in the awkward situation of
having a potential explanation for the small Yukawa couplings of e.g. the down-type quarks,
but not the up-type quarks.  It is clearly more desirable to build fully composite generations.
Here we describe theories that produce one or two composite generations.

\subsection{Producing one full generation in the ${\bf 10} + {\bf \bar 5}$}

To produce one full composite generation, we consider SQCD with $N_c=11$ and $N_f=12$. The $SU(12)$ global symmetry is higgsed down to $SU(11)$ by the dual quark vev in \eref{metastable_vacuum}. The MSSM gauge group is an $SU(5)$ gauged subgroup of $SU(11)$. We choose the following embedding of $SU(5)$ into $SU(12)$  

\beq
\begin{array}{rcl}
Q & = & ({\bf 5}+ {\bf \overline{5}} + {\bf 1})+ {\bf 1}\\
\tilde{Q} & = & ({\bf \overline{5}} + {\bf 5} + {\bf 1})+ {\bf 1}
\end{array}
\label{embedding_12}
\eeq
where the decomposition in parenthesis indicates the embedding into $SU(11)$. The mesons of the magnetic theory can be decomposed according to
\beq
\Phi=\left(\begin{array}{ccc} Y_{1\times 1} & & Z^T_{1\times 11} \\
\tilde{Z}_{11\times 1} & & \Phi_{0,11\times 11} \end{array}\right)
\eeq
with $Y$, $Z$, $\tilde{Z}$ and $\Phi_0$ transforming in the ${\bf 1}$, ${\bf \overline{11}}$, ${\bf 11}$ and $(\rm{{\bf Adj}}+{\bf 1})$ of $SU(11)$. $\Phi_0$ decomposes into

\beq
\Phi_0 = ({\bf 10}+{\bf \bar{5}})+\left[2 \times {\bf 24} + {\bf 15} + {\bf \overline{15}}+{\bf \overline{10}}+ 2\times {\bf 5}+ {\bf \overline{5}}+3 \times {\bf 1}\right]
\label{Phi_0_decomposition}
\eeq
under $SU(5)$, where we identify the desired composite generation plus additional matter. In addition to the matter content of the ISS model, we introduce two elementary $({\bf 10}+{\bf \bar{5}})$ generations, which we denote $\Psi^{(1)}$ and $\Psi^{(2)}$, and which correspond to two elementary Standard Model generations. And, as in \S3, we introduce spectator fields to the SQCD dynamics which are charged
under $SU(5)$, one for each extra charged matter field in (\ref{Phi_0_decomposition}). 
This means we have fields $X_{\bf 24}, X'_{\bf 24}, X_{\bf 15}, X_{\bf \overline {15}}, X_{\bf 10 + \bar 5},
X'_{\bf \bar 5}, X'_{\bf 5}$. (One can also remove the singlets other than ${\rm Tr}(\Phi_0)$ in this
manner, if one chooses).

In order to give masses to the additional matter that arises as mesons, we introduce the following superpotential couplings in the electric theory 

\beq
W_3=  \lambda \sum_{\bf \overline{R}} \left((Q \tilde Q)_{\bf R} X_{\bf \overline{R}}\right)
\label{W3}
\eeq
where ${\bf \overline{R}}$ runs over all of the $SU(5)$-representations of the $X$ fields, and the Yukawa coupling pairs them up with the unwanted components of the meson. 
These couplings give masses $\sim {\Lambda}_{e}$ to all extra charges arising from $\Phi_0$, leaving
just the composite generation light.
We also add the quartic coupling in \eref{quartic} to break the R-symmetry.

In summary, the matter content of the electric theory is
\beq
\begin{array}{c|ccc}
 & \ \ \ SU(11)_{\rm gauge} \ \ \ & & SU(5)_{\rm SM} \subset SU(11)_{\rm global} \\ \hline
\ \ \ Q \ \ \ & {\bf 11} & & {\bf 5}+ {\bf \overline{5}} + {\bf 1} + {\bf 1}\\
\ \ \ \tilde{Q} \ \ \ & {\bf \overline{11}} & & {\bf \overline{5}} +{\bf 5} + {\bf 1} + {\bf 1}\\
\Psi^{(1,2)} & {\bf 1} & & {\bf 10}+{\bf \bar{5}} \\
X_{\bf 24} & {\bf 1} & & {\bf 24}\\
X'_{\bf 24} & {\bf 1} & & {\bf 24}\\
X_{\bf 15} & {\bf 1} & & {\bf 15}\\
X_{\bf \overline{15}} & {\bf 1} & & {\bf \overline{15}}\\
X_{\bf 10 } & {\bf 1} & & {\bf  10}\\
X_{\bf \bar 5} & {\bf 1} & & {\bf \bar 5}\\
X'_{\bf \bar 5} & {\bf 1} & & {\bf \bar 5}\\
X'_{\bf 5} & {\bf 1} & & {\bf 5}
\end{array}
\label{matter_electric}
\eeq

The matter in the magnetic theory is given by

\beq
\begin{array}{cc}

\begin{array}{c}
\\ \Phi \ \ \left\{\begin{array}{c}  \\ \\ \\ \\ \\ \end{array}\right.
\end{array} &

\begin{array}{c|c}
 & \ \ \ SU(5)_{\rm SM} \subset SU(11)_{\rm global} \ \ \ \\ \hline
\ \ \ q \ \ \ & {\bf \overline{5}} +{\bf 5} + {\bf 1} + {\bf 1} \\
\ \ \ \tilde{q} \ \ \ & {\bf 5}+ {\bf \overline{5}} + {\bf 1} + {\bf 1} \\ \hline
Y & {\bf 1} \\
Z & {\bf \overline{5}} +{\bf 5} + {\bf 1} \\
\tilde{Z} & {\bf 5}+ {\bf \overline{5}} + {\bf 1} \\
\Phi_0 & \ \ \ \begin{array}{rr}({\bf 10}+{\bf \bar{5}})+ [2\times {\bf 24} + {\bf 15} + {\bf \overline{15}}+ \\  +{\bf \overline{10}}+2 \times {\bf 5}+ {\bf \overline{5}}+3 \times {\bf 1} ]\end{array} \ \ \ \\ \hline
\Psi^{(1,2)} &  {\bf 10}+{\bf \bar{5}} \\
X_{\bf R} & {\bf R} = 2\times {\bf 24} + {\bf 15} + {\bf \overline{15}} + {\bf 10}+ {\bf 5} +2 \times {\bf \bar{5}} 
\end{array}
\end{array}
\label{matter_magnetic}
\eeq
To save space, we have compressed the spectator representations into a single line in (\ref{matter_magnetic}); the spectators
to the SQCD dynamics are enumerated individually in (\ref{matter_electric}), and we trust this will cause no confusion.

In addition to the light fields we have already enumerated above, one should keep in mind that there are additional light ``messengers'' from the $(\rho,Z)$ sector. These transform in the 4 $\times \left({\bf 5 + \bar 5}\right)$ of $SU(5)$.

\subsection{Models with two composite generations}

To produce a model with 2 composite generations (we could further extend the model to accommodate three, but the large value
of the top-quark Yukawa coupling makes one hesitant to pursue that interpretation), we start
with SQCD with $N_c = 16$ and $N_f = 17$.  The $SU(17)$ global symmetry is
higgsed down to $SU(16)$ by the quark condensate in the ISS metastable vacuum.  The MSSM gauge group
is a weakly gauged $SU(5)$ subgroup of $SU(16)$; the embedding is such that, in the electric
theory, the quark fields transform as

\beq
\begin{array}{rcl}
Q & = & ({\bf 5}+ {\bf 5} + {\bf \overline{5}} + {\bf 1})+ {\bf 1}\\
\tilde{Q} & = & ({\bf \overline{5}} + {\bf \overline{5}} + {\bf 5} + {\bf 1})+ {\bf 1}
\end{array}
\eeq
where the parenthesis separate the ${\bf 16}$ and ${\bf \overline{16}}$ of $SU(16)$ from the
singlet.  The magnetic mesons of the theory take the form

\beq
\Phi =\left(\begin{array}{ccc} Y_{1\times 1} & & Z^T_{1\times 16} \\
\tilde{Z}_{16\times 1} & & \Phi_{0,16\times 16} \end{array}\right)
\eeq
with $Y, Z, \tilde Z$ and $\Phi_0$ transforming in the ${\bf 1}$, ${\bf \overline{16}}$, ${\bf 16}$ and $({\rm {\bf Adj}}+{\bf 1})$ of $SU(16)$.  $\Phi_0$ decomposes into $SU(5)$ representations as:

\beq
\Phi_0 = 2 \times ({\bf 10 + \bar 5}) + \left[5 \times {\bf 24} + 2 \times {\bf 15} + 2 \times {\bf \overline{15}} +
2 \times {\bf \overline{10}} + 3 \times {\bf 5} + {\bf \bar 5} + 6 \times {\bf 1}\right]~.
\label{Phimess}
\eeq
We have put the two composite generations as the first contributions on the right hand side of
(\ref{Phimess}).  All of the other charged matter fields in $\Phi_0$, we can lift as in all of the previous models: we add appropriate spectators $X_{\bf R}$ to the SQCD dynamics which transform under
the global (soon to be weakly gauged) SU(5), and we pair them up via the analogue of
(\ref{crapmass_1}).  
These states can then be integrated out at the high scale $\Lambda_{e}$.
In addition, we need to add an ``elementary" third MSSM generation 
$\Psi^{(3)}$ in the $({\bf 10} + {\bf \bar 5})$ of $SU(5)$.

As in \S4.1, in addition to these states, there are  $6 \times \left( {\bf 5 + \bar 5}\right)$ ``messengers'' coming from the $(\rho,Z)$ sector.

\section{Models with fewer un-necessary very massive states at ${\Lambda}_{\rm electric}$}

All models discussed so far embed the MSSM composite fields into the adjoint of an $SU(N_c)$ subgroup of the global symmetry group. As a result, a large number of extra fields charged under the MSSM are generated. While we have explained how to give these fields large masses via cubic superpotential couplings of the form \eref{W3}, this abundance of extra matter is to some extent aesthetically unpleasant.

A simple way to improve this situation is to consider other gauge theories with metastable vacua in which the magnetic mesons (more concretely the pseudomoduli analogous to $\Phi_0$) appear in smaller representations. SQCD with $Sp(N_c)$ gauge group and massive flavors is a natural candidate for such a construction. In this case, the size of $\Phi_0$ is roughly reduced by a factor two. We now discuss this model very briefly, focusing on the reduction of the additional matter content. We refer the reader to \cite{Intriligator:1995ne} and \cite{ISS} for details on Seiberg duality and metastable SUSY breaking in these theories.

Let us consider $Sp(N_c)$ SQCD with $N_f$ flavors, given by the fields $Q_i$ ($i=1,\ldots,2N_f$) in the fundamental representation. The Seiberg dual theory has  $Sp(N)$ gauge group ($N=N_f-N_c-2$), $N_f$ flavors of dual quarks $q_i$ ($i=1,\ldots,2N_f$), mesons $\Phi$ and a cubic superpotential coupling dual quarks and mesons \cite{Intriligator:1995ne}. The free-magnetic range corresponds to $N_c+3 \leq N_f  \leq {3 \over 2} (N_c+1)$. The global symmetry is

\beq
SU(2 N_f) \times U(1)' \times U(1)_R \, .
\eeq
The $q_i$ and $\Phi$ transform in the ${\bf\overline{2N_f}}$ and antisymmetric representations of $SU(2N_f)$, respectively. The ISS construction extends to this model, i.e. if we add a mass term for the quarks in the electric theory, SUSY is broken in a mestastable vacuum. Adding an equal non-vanishing mass to all flavors of the electric theory, the global symmetry is reduced to 

\beq
Sp(N_f) \times U(1)_R
\eeq
At the metastable vacuum, the dual quarks acquire a vev, breaking the global symmetry down to

\beq
Sp(N)_D \times Sp(N_c+2) \times U(1)_R
\eeq
In complete analogy with the $SU(N_c)$ case, the meson matrix contains pseudomoduli parametrized by a $2(N_c+2) \times 2(N_c+2)$ antisymmetric sub-matrix $\Phi_0$ inside $\Phi$, in which we will embed the MSSM composites.

Now, let us construct a model with one full composite generation. Consider $Sp(4)$ gauge theory with $N_f=7$. As in previous examples, the magnetic gauge group is trivial. 
The global symmetry group of this theory includes an $SU(14)$ factor which is spontaneously
broken to $Sp(6)$ in the metastable vacuum.  We embed $SU(5)$ into $SU(14)$ so that
\beq
\begin{array}{rcl}
Q & = & {\bf 5} + {\bf \overline{5}} + 4 \times {\bf 1}~.\\
\end{array}
\eeq

The dual meson transforms in the ${\bf  91}$ of $SU(14)$, and the pseudomoduli ${\bf \Phi_0}$
transform in the ${\bf 65} + {\bf 1}$ of $Sp(6)$ (with the first being the traceless anti-symmetric
representation).  In terms of $SU(5)$ quantum numbers, one has
\beq
\Phi_0 = ({\bf 10} + {\bf \overline 5}) + \left[ {\bf 24} + {\bf \overline {10}} + 2 \times {\bf 5} + {\bf \overline 5} + 2 \times {\bf 1}\right]~.
\eeq
This model therefore accommodates one full composite generation, with less extra
$SU(5)$-charged matter than the theory in \S4.1.

\section{Brief comments on phenomenology}

We close with some very elementary remarks about the phenomenology of the most
complete "single-sector" models we can generate in this way, which have two full composite generations. The analysis of other models is completely analogous.

\bigskip
\noindent
$\bullet$
As we have already emphasized, the main virtue of our models is their calculability. It is straightforward to compute their spectrum, and we have already discussed most of it in previous sections. The mass scales for chiral multiplets charged under the MSSM gauge group can be summarized in the following table, which gives the masses in the limit of ${\it vanishing}$ Standard Model gauge couplings (and for the moment ignoring the fermion masses which arise after $SU(2) \times U(1)$ breaking) :

\beq
\begin{array}{ccccccc}
\hline
& & & \ \ & {\rm Fermions} & \ \ \ & {\rm Bosons} \\ \hline 
\ \ {\rm MSSM} \ \ && \ \ \Psi^{(1,2)} & & 0 & & h^2 \mu \\
 & & \ \ \Psi^{(3)} & & 0 & & 0 \\ \hline
\ \ {\rm Messengers} \ \ & & \ \ {\bf R},{\bf \overline{R}} & & \Lambda_e & & \Lambda_e \\
 & & \ \ \rho, \tilde{\rho}, Z, \tilde{Z} & & h \mu  & & h \mu \\
 & & & & h \mu  & & 0 \\
\hline
\end{array}
\label{spectrum}
\eeq

\bigskip

\noindent 
The last line corresponds to the Nambu-Goldstone bosons associated with the $SU(N_f)\to SU(N)\times SU(N_c)$ breaking of the global symmetry. They acquire 
a mass $\sim g_{\rm SM} \mu$ once we gauge the MSSM gauge group. In the ``messengers" category we collect all additional massive matter charged under the MSSM. The reason for this classification is that these fields (together with the composite MSSM fields) gauge mediate SUSY breaking. The leading contribution comes from the $(\rho,Z)$ sector, which is much lighter than the ${\bf R}$ and ${\bf \overline{R}}$ fields. The gauge mediation contribution to sfermion masses (which is the only contribution in the case of third generation scalars) is $\sim g_{\rm SM}^2 \mu$ \cite{Strassler}.

The MSSM gaugino gets a mass of order $g_{\rm SM}^2 \mu_\phi$, with $\mu_\phi$ the R-symmetry breaking parameter. In addition to the fields listed in \eref{spectrum}, we have the MSSM singlets $\Phi_{0,{\bf 1}}$, $Y$, $\chi$ and $\tilde{\chi}$. $\Phi_{0,{\bf 1}}$ is lifted by a Coleman-Weinberg potential as discussed in the appendix. $Y$ and $(\chi+\tilde{\chi})$ get masses $\sim h \mu$. We can give a mass of order $g_B \mu$ to $(\chi-\tilde{\chi})$ by gauging $U(1)_B$.\footnote{Notice that all our examples have a trivial magnetic gauge group (i.e. $N=1$) and hence $Y$, $\chi$ and $\tilde{\chi}$ are just $1\times 1$ dimensional.} We refer the reader to \cite{Strassler} for details. 

\bigskip

\noindent
$\bullet$ In our models, the first two generations arise as composite dimension two operators
in the UV of the electric theory -- these are ``meson models," in the parlance of 
\cite{LT}.  Therefore, we can hope to explain some but not all of the structure of the 
Yukawa couplings that is evident in the Standard Model.  Assuming that the Higgs particles
are added as elementary states relative to the strongly coupled dynamics, the Yukawa couplings
of the first two generations
(neglecting mixing with the third generation) arise from dimension 6 operators in the high-energy theory,
and are naturally suppressed.  More precisely, one expects the Yukawa couplings to be generated
by a superpotential of the schematic form
\be
{\cal W}_{\rm Yuk} \sim {1\over M_{\rm flavor}^{2}} (Q \tilde Q) H (Q \tilde Q) +
{1 \over M_{\rm flavor}} (Q\tilde Q) H \Psi^{(3)} + \Psi^{(3)} H \Psi^{(3)}~.
\ee
This gives rise to a Yukawa matrix whose basic structure is
\be
\begin{pmatrix}
\epsilon^2 ~& \epsilon^2~ & \epsilon\\
\epsilon^2~ & \epsilon^2~ & \epsilon\\
\epsilon~ & \epsilon~ & 1
\end{pmatrix}
\ee
controlled by the small parameter $\epsilon \sim {\Lambda_{e} \over M_{\rm flavor}}$.  For
$\epsilon \sim 10^{-2}$, this is a reasonable starting point for obtaining the correct Yukawa couplings,
though some additional structure remains to be explained.

Clearly, one could do a better job of explaining the flavor structure visible in the Standard Model with a composite model where the first generation arises from a composite operator which has (a lowest
component in its supermultiplet of) dimension three in the high energy theory, while the second generation comes from an operator of dimension two.  It would be interesting to construct such models following our general strategy.  The metastable vacua discussed
by \cite{Amariti}, which arise in supersymmetric QCD with additional adjoint matter fields, would 
likely be a good starting point, since the magnetic duals incorporate additional mesons which are
cubic in the electric variables.

\bigskip

\noindent
$\bullet$ The MSSM gauginos and the third generation elementary states get their soft masses, in these
models, from gauge mediation arising from both the composite MSSM generations, and from the
massive messengers (additional vector-like MSSM charges) which exist at the scale $\sim h\mu$.
For this reason, it is desirable to have the first two generations of sparticles (and the messengers)
at  a mass scale somewhat above the TeV scale. For $h \sim 1$, we should choose
$\mu \sim {\rm 5 ~TeV}$.  The resulting stop mass will be small enough to avoid excessive
fine-tuning; the resulting large masses for the squarks in the first two generations will help with
FCNCs, as in the scenarios of \cite{Dine,Giudice,Pomarol,Nelson}.  It is interesting to consider fully solving
the supersymmetric flavor problem by raising the masses of the first two generation sparticles
a bit more - for discussions of whether this is possible or not, see also \cite{Nima,Agashe,Nomuratwo,Nomura}.

\bigskip

\noindent
$\bullet$ With $\mu \sim {\rm 5 ~TeV}$, we are still allowed great freedom in our choice of the compositeness scale
$\Lambda_{e}$, since $\mu \sim \sqrt{m \Lambda_{e}}$.  To avoid Landau poles before the
GUT scale, it is clearly desirable to choose $\Lambda_{e}$ as large as possible.  Even for
very large $\Lambda_{e}$ (close to the GUT scale), the scenario of \S4.2\ will suffer from Landau poles at or before
the GUT scale, due to the large number ($ 6 \times ~({\bf 5 + \bar 5})$)
of additional messengers which make a direct
mediation contribution to soft masses and which appear with masses $\sim {\rm 5 ~TeV}$.  
One way to avoid this problem may be to consider more general gauge groups as in \S5.
In any event, clearly the proliferation of extra $SU(5)$ charges at a high scale is the least attractive feature of the simple models we have proposed here, and finding more economical models would be a natural goal for future research.  On the
other hand, it is fortunate that we are at least free to push $\Lambda_e$ to high scales to avoid many
exotics anywhere close to the TeV scale.  For a general discussion of the issue of Landau poles
in a similar context, see \cite{Khoze}.

\bigskip

\noindent
$\bullet$ Single-sector models typically involve strong coupling, both in the 
SUSY-breaking dynamics and in the formation of the composite MSSM states.  For this reason it
has been difficult to exhibit calculable examples.   We did so here by making use of
the electric/magnetic duality of supersymmetric gauge theories \cite{SeibergDual}.   Another
tool which could be useful in this regard is gauge/gravity duality.  For discussions of how one might approach the problem from this perspective, see \cite{Gherghetta, us, toappear}.

\bigskip

\subsubsection*{Note Added:}  
After this paper was published, we learned that the precise models studied here admit other
metastable vacua which are lower in energy than the vacua we focused on.  Then, a more detailed study 
is required to determine whether the lifetime of our vacua are sufficient to accommodate realistic cosmology.  
However, a very minor change to the models -- adjusting some of the electric quark masses by 
an $\mathcal{O}(1)$ factor -- removes this issue and leaves the rest of our
discussion unchanged.  For details, see \S4 of \cite{Behbahani:2010wh}.
We thank D. Green, A. Katz and Z. Komargodski
for bringing this issue to our attention.

\bigskip
\bigskip
\centerline{\bf{Acknowledgements}}
\medskip
We would like to thank F. Benini, A. Dymarsky, D. Simic and H. Verlinde for enjoyable collaborations
on related issues, using the different technique of gauge/gravity duality. S. F. would like to thank G. Torroba for very interesting discussions. S. K. also thanks S. Dimopoulos, T. Gherghetta and J. Wacker for enjoyable discussions in 2007 which 
stimulated his interest in this subject.  S.K. acknowledges the hospitality of the
Kavli Institute for Theoretical Physics and the Aspen Center for Physics while this work was
in progress.  S.K. was supported in part by the Stanford Institute for Theoretical Physics,
the NSF under grant PHY-0244728, and the DOE under contract DE-AC03-76SF00515.
S.F. and S.K. both acknowledge the support of the National Science Foundation 
under Grant No. PHY05-51164.

\newpage

\appendix

\section{SUSY-breaking vacua after adding cubic and quartic couplings in the electric theory}

\label{app: cubics}

In this appendix we study the effect of adding the superpotential terms \eref{crapmass_1} and \eref{quartic} to the electric theory.

\subsection*{Cubic couplings}

Let us first investigate what happens when adding \eref{crapmass_1}. Our main goal is to see that the ISS metastable vacuum is not destabilized as long as the coupling does not involve ${\rm Tr} ~\Phi_0$. The $\Phi_0$ matrix can be decomposed into an adjoint and a singlet of $SU(N_c)$

\beq
\Phi_0=\Phi_{0,{\rm {\bf Adj}}}+\Phi_{0, {\bf 1}} \, .
\eeq
$\Phi_{0,{\rm {\bf Adj}}}$ is traceless and hence can be parametrized as a general $N_c \times N_c$ matrix subject to the constraint
\beq
\Phi_{0,{\rm {\bf Adj}},N_c}^{N_c}=-\sum_{a=1}^{N_c-1}\Phi_{0,{\rm {\bf Adj}},a}^a~.
\eeq 
The singlet is proportional to the identity matrix
\beq
\Phi_{0, {\bf 1}}={1\over N_c}({\rm Tr} \Phi_0) {\bf 1}_{N_c}~.
\eeq

For our purposes, it is sufficient to consider the effect of coupling the entire adjoint $\Phi_{0,{\rm {\bf Adj}}}$ to an adjoint field $X$; we will briefly discuss cases where we pair up less of $\Phi_{0,{\rm{\bf Adj}}}$ at
the end. The magnetic superpotential is

\beq
W=h \tilde{q}_i \Phi^i_j q^j - h \mu^2 \Phi^i_i +\lambda \Phi^i_{0,{\rm {\bf Adj}},j} X^j_i
\label{W_mag}
\eeq
Let us define indices $\mu,\nu=1,\ldots,N$ and $a,b=1,\ldots,N_c$.

Since $X$ transforms in the adjoint representation of $SU(N_c)$, it is traceless, i.e. $X^{N_c}_{N_c}=-\sum_{a=1}^{N_c-1} X^a_a$. We can then rewrite the last term in \eref{W_mag} in terms of $\Phi$

\beq
{\rm Tr} [\Phi_{0,{\rm {\bf Adj}}} \, X]={\rm Tr} [(\Phi_{0,{\rm {\bf Adj}}}+\Phi_{0, {\bf 1}}) \, X]={\rm Tr} [\Phi_0 \, X]={\rm Tr} [\Phi \, X]
\eeq
where we have used $\Phi_{0, {\bf 1}} X\sim X$, and the fact that $X$ is traceless. 
In the last equality, we define ${\rm Tr}[\Phi~ X]$ by extending $X$ to be an $N_f \times N_f$ matrix,
which is identically zero except in the $N_c \times N_c$ block that we really need for our spectators (so all the additional spurious contributions in the ${\rm Tr}$, coming from the matrix entries with one or both indices running over $\mu,\nu$ on $\Phi$ and $X$, vanish identically).

This rewriting simplifies the analysis of F-terms -- we only have to keep track of the tracelessness of $X$, without taking special care of the decomposition of $\Phi_0$. The F-terms become:

\beq
F_{\Phi^{\mu}_{\nu}}={\partial W \over \partial \Phi^{\mu}_{\nu}}=h \tilde{q}_\mu q^\nu -h \mu^2 \delta_{\mu}^{\nu}
\label{F_Y}
\eeq

\beq
F_{\Phi^{N+a}_{N+b}}={\partial W \over \partial \Phi^{N+a}_{N+b}}={\partial W \over \partial \Phi^a_{0,b}}=h \tilde{q}_{N+a} q^{N+b} -h \mu^2 \delta^{a}_b + \lambda X^a_b
\label{F_Phi0}
\eeq

\beq
F_{\Phi^{N+a}_{\nu}} = {\partial W \over \partial \Phi^{N+a}_{\nu}} = h \tilde q_{N+a} q^{\nu}
\label{newone}
\eeq

\beq
F_{\Phi^{\mu}_{N+b}} = {\partial W \over \partial \Phi^{\mu}_{N+b}} = h \tilde q_{\mu} q^{N+b}
\label{newtwo}
\eeq

\beq
F_{X^a_b}={\partial W \over \partial X^a_b}=\lambda \Phi^a_b- \lambda \delta^{a}_{b} \Phi^{N_c}_{N_c}\ \  \ \ a,b = 1,\cdots, N_c ~ ({\rm but~ not ~both}~ N_c) 
\label{F_X}
\eeq

\beq
F_{q^j}={\partial W \over \partial q^j}= h \tilde{q}_i \Phi^i_j
\label{F_q}
\eeq

\beq
F_{\tilde{q}_j}={\partial W \over \partial \tilde{q}_i}= h  \Phi^j_i q^i
\label{F_qt}
\eeq
where there is no sum over repeated $\mu$ and $a$ indices. The $(\tilde q q)$ matrix has at most rank $N$. Since the F-terms for the $\Phi$
fields cannot all vanish, this model gives rise to SUSY breaking by a generalization of the rank condition in \cite{ISS}.  

A local minimum of the scalar potential is found as follows. From equation \eref{F_Y}, we can see
that $F_{\Phi^{\mu}_{\nu}}$ vanishes if one only turns on the $\chi$, $\tilde \chi$ components of
the magnetic squarks:
\beq
\langle\chi\rangle = q_0 {\bf 1}_{N\times N},~\langle \tilde \chi \rangle = \tilde q_0 {\bf 1}_{N \times N},~
\tilde q_0 q_0 = \mu^2 \rightarrow 
\langle \tilde q q \rangle={\rm diag}(\mu^2 {\bf 1}_N,0_{N_c})~.
\eeq
 Plugging this into \eref{F_Phi0}, together with ${\rm Tr} X=0$, sets $X=0$.  The tracelessness constraint is crucial -- in its absence, one would instead find only supersymmetric vacua where $X$ condenses and breaks the Standard
Model gauge group.   The $\rho, \tilde \rho$ components of the magnetic squarks get a positive mass$^2$ from
their coupling to the non-zero $F_{\Phi_0}$.

The F-components in \eref{newone} and \eref{newtwo} vanish with this choice of quark vevs.
The F-components of the magnetic quarks \eref{F_q} and \eref{F_qt} vanish for $\Phi={\rm diag}(0_N,\Phi_0)$, with arbitrary $\Phi_0$. 
Finally, \eref{F_X} vanishes for $\Phi_{0,{\rm {\bf Adj}}}=0$ and $\Phi_{0,{\bf 1}}$ arbitrary.

Expanding around these background vacuum expectation values, the only non-zero F-components
arise from \eref{F_Phi0}:
\beq
F_{\Phi^{N+a}_{N+b}} = - h\mu^2 \delta^{a}_{b}~.
\ee
A significant difference, as compared to the analysis in \cite{ISS}, is that most of the $\Phi_0$ degrees
of freedom are required to vanish at tree-level due to the $X$ F-term.  This is not a surprise; we have
added the spectators $X$ precisely to give these components a large tree-level mass.

\bigskip

\subsection*{Including the quartic coupling}

It remains to discuss the stabilization of $\Phi_{0,\bf{1}}$.   
We now include the coupling (\ref{quartic}) which breaks R-symmetry
\beq
W \to W + W_4, ~~W_4 = h^2\mu_{\phi}tr(\Phi^2)~.
\eeq
As in the pure ISS model, only $\rho$, $\tilde \rho$, $Z$ and $\tilde Z$ have a non-supersymmetric mass matrix at tree-level and hence contribute
to the effective Coleman-Weinberg potential. Furthermore, our couplings in this sector are identical to the ones considered in \cite{Strassler}. Hence, we can proceed exactly as in its appendix; we include a brief discussion here only for completeness. The fermionic and bosonic messenger masses are
\beq
\begin{array}{rcl}
m^2(\Phi_{0,\bf{1}}) & = & |h|^2 [ ~|q_0|^2 + {1\over 2}|\Phi_{0,{\bf 1}}|^2 + {1\over 2} |h\mu_{\phi}|^2 \\
 & & + {1\over 2} \sigma \sqrt{(|\Phi_{0,\bf{1}}|^2 - |h\mu_{\phi}|^2)^2 + 4 |q_0 \Phi_{0,{\bf 1}}^* +
 q_0^* h\mu_{\phi}|^2} ~]
\end{array}
\eeq
\beq
\begin{array}{rcl}
\tilde m^2 (\Phi_{0,{\bf{1}}}) & = & |h|^2 [~ |q_0|^2 + {1\over 2} |\Phi_{0,{\bf 1}}|^2 + {1\over 2} |h\mu_{\phi}|^2
+ {1\over 2}\eta |\mu^2 - h\mu_{\phi}\Phi_{0,\bf{1}}\vert \\
& &+ {1\over 2} \sigma \sqrt{(|\Phi_{0,\bf{1}}|^2 - |h\mu_{\phi}|^2 + \eta|\mu^2 - h\mu_{\phi}
\Phi_{0,\bf{1}}|)^2 + 4 |q_0 \Phi_{0,\bf 1}^* + q_0^* h\mu_{\phi}|^2}~]
\end{array}
\eeq
where, for simplicity, we only consider the dependence on $\Phi_{0,{\bf 1}}$ and set all other pseudomoduli to their stabilized values. Here $\sigma$, $\eta$ run over $\pm 1$.  There are 4 $N_c N$ fermions and 2 $N_c N$ complex bosons.

The one-loop Coleman-Weinberg potential for $\Phi_{0,{\bf 1}}$ is then easily determined by integrating
out the messenger fields.  The result is as in \cite{Strassler}; the $\Phi_{0,\bf{1}}$ field has a stable
vacuum (for $\mu_{\phi} \ll \mu$), resulting from a competition between the tree-level potential and the
1-loop corrections, 
 with a vacuum expectation value
\beq
h\Phi_{0,{\bf 1}} \sim {\mu^2 \mu_{\phi}^* \over {b|\mu|^2 + |\mu_{\phi}|^2}}
\eeq
and a 
mass of order 
\beq
m^2_{\Phi_{0,\bf{1}}} ~\sim~{({\rm log}~{\rm 4} - 1)\over 8\pi^2} N |h^4 \mu^2| + |h^4 \mu_{\phi}^2|~.
\eeq

It is easy to discuss also the cases where one only decouples some subset of the 
$\Phi_{0,{\bf Adj}}$ degrees of freedom by pairing them with spectator $X$ fields.  Instead of being locked at zero classically by a large ${\cal O}(\Lambda_e)$ mass, the unpaired components of $\Phi_{0,{\bf Adj}}$ will be given a tiny mass$^2$ $\sim h^4 \mu_{\phi}^2$ at tree level.  As discussed in \cite{Strassler}, in the full theory,
the more significant effect is that
the bosonic pieces of the unpaired supermultiplets will get a positive mass$^2$ of ${\cal O}(h^4 \mu^2)$
at one loop (which is a much larger effect because we work in the regime $\mu \gg \mu_{\phi}$).   In the fermionic sector, as discussed in \S3.1, one will get light composite fermions (which are
predominantly from the unpaired fermions in $\Phi_0$ for $\Lambda_e \gg h^2\mu_{\phi}$, though they contain some small admixture of the external
$X$ fermions with the same quantum numbers).

\newpage


\begin{thebibliography}{100}

\bibitem{ALT}
N.~Arkani-Hamed, M.~A.~Luty and J.~Terning,
  ``Composite quarks and leptons from dynamical supersymmetry breaking  without
  messengers,''
  Phys.\ Rev.\  D {\bf 58}, 015004 (1998)
  [arXiv:hep-ph/9712389].

\bibitem{LT}
M.~A.~Luty and J.~Terning,
  ``Improved single sector supersymmetry breaking,''
  Phys.\ Rev.\  D {\bf 62}, 075006 (2000)
  [arXiv:hep-ph/9812290].

\bibitem{Dine}
M.~Dine, A.~Kagan and S.~Samuel,
  ``Naturalness in supersymmetry, or raising the supersymmetry breaking scale,''
  Phys.\ Lett.\  B {\bf 243}, 250 (1990).

\bibitem{Giudice}
S.~Dimopoulos and G.~F.~Giudice,
  ``Naturalness constraints in supersymmetric theories with nonuniversal soft
  terms,''
  Phys.\ Lett.\  B {\bf 357}, 573 (1995)
  [arXiv:hep-ph/9507282].

\bibitem{Pomarol}
A.~Pomarol and D.~Tommasini,
  ``Horizontal symmetries for the supersymmetric flavor problem,''
  Nucl.\ Phys.\  B {\bf 466}, 3 (1996)
  [arXiv:hep-ph/9507462].

\bibitem{Nelson}
A.~G.~Cohen, D.~B.~Kaplan and A.~E.~Nelson,
  ``The more minimal supersymmetric standard model,''
  Phys.\ Lett.\  B {\bf 388}, 588 (1996)
  [arXiv:hep-ph/9607394].

 \bibitem{DG}
  S.~Dimopoulos and H.~Georgi,
  ``Softly Broken Supersymmetry And SU(5),''
  Nucl.\ Phys.\  B {\bf 193}, 150 (1981).

\bibitem{ISS}
K.~A.~Intriligator, N.~Seiberg and D.~Shih,
  ``Dynamical SUSY breaking in meta-stable vacua,''
  JHEP {\bf 0604}, 021 (2006)
  [arXiv:hep-th/0602239].

\bibitem{SeibergDual}
 N.~Seiberg,
  ``Electric - magnetic duality in supersymmetric nonAbelian gauge theories,''
  Nucl.\ Phys.\  B {\bf 435}, 129 (1995)
  [arXiv:hep-th/9411149].
  
\bibitem{Strassler}
  R.~Essig, J.~F.~Fortin, K.~Sinha, G.~Torroba and M.~J.~Strassler,
  ``Metastable supersymmetry breaking and multitrace deformations of SQCD,''
  JHEP {\bf 0903}, 043 (2009)
  [arXiv:0812.3213 [hep-th]].
   
\bibitem{Torroba}
 R.~Essig, K.~Sinha and G.~Torroba,
  ``Meta-Stable Dynamical Supersymmetry Breaking Near Points of Enhanced
  Symmetry,''
  JHEP {\bf 0709}, 032 (2007)
  [arXiv:0707.0007 [hep-th]].
   
\bibitem{Kutasov}
  A.~Giveon and D.~Kutasov,
  ``Stable and Metastable Vacua in Brane Constructions of SQCD,''
  JHEP {\bf 0802}, 038 (2008)
  [arXiv:0710.1833 [hep-th]].

\bibitem{Intriligator:1995ne}
  K.~A.~Intriligator and P.~Pouliot,
  ``Exact superpotentials, quantum vacua and duality in supersymmetric SP(N(c))
  gauge theories,''
  Phys.\ Lett.\  B {\bf 353}, 471 (1995)
  [arXiv:hep-th/9505006].

\bibitem{Amariti}
A.~Amariti, L.~Girardello and A.~Mariotti,
  ``Non-supersymmetric meta-stable vacua in SU(N) SQCD with adjoint matter,''
  JHEP {\bf 0612}, 058 (2006)
  [arXiv:hep-th/0608063].

\bibitem{Nima}
  N.~Arkani-Hamed and H.~Murayama,
  ``Can the supersymmetric flavor problem decouple?,''
  Phys.\ Rev.\  D {\bf 56}, 6733 (1997)
  [arXiv:hep-ph/9703259].

\bibitem{Agashe}
K.~Agashe and M.~Graesser,
  ``Supersymmetry breaking and the supersymmetric flavour problem: An  analysis
  of decoupling the first two generation scalars,''
  Phys.\ Rev.\  D {\bf 59}, 015007 (1999)
  [arXiv:hep-ph/9801446].

\bibitem{Nomuratwo}
  J.~Hisano, K.~Kurosawa and Y.~Nomura,
  ``Large squark and slepton masses for the first-two generations in the
  anomalous U(1) SUSY breaking models,''
  Phys.\ Lett.\  B {\bf 445}, 316 (1999)
  [arXiv:hep-ph/9810411].

\bibitem{Nomura}
 Y.~Nomura,
  ``Decoupling solution to the supersymmetric flavor problem without color
  instability,''
  arXiv:hep-ph/9909281.

\bibitem{Khoze}
 S.~Abel and V.~V.~Khoze,
  ``Direct Mediation, Duality and Unification,''
  JHEP {\bf 0811}, 024 (2008)
  [arXiv:0809.5262 [hep-ph]].
   
\bibitem{Gherghetta}
M.~Gabella, T.~Gherghetta and J.~Giedt,
  ``A gravity dual and LHC study of single-sector supersymmetry breaking,''
  Phys.\ Rev.\  D {\bf 76}, 055001 (2007)
  [arXiv:0704.3571 [hep-ph]].

\bibitem{us}
 F.~Benini, A.~Dymarsky, S.~Franco, S.~Kachru, D.~Simic and H.~Verlinde,
  ``Holographic Gauge Mediation,''
  arXiv:0903.0619 [hep-th].

\bibitem{toappear}
F.~Benini, A.~Dymarsky, S.~Franco, S.~Kachru, D.~Simic and H.~Verlinde,
to appear.

\bibitem{Behbahani:2010wh}
  S.~R.~Behbahani, N.~Craig and G.~Torroba,
  ``Single-sector supersymmetry breaking, chirality, and unification,''
  arXiv:1009.2088 [hep-ph].

\end{thebibliography}
\end{document}